\documentclass[reviewcopy]{elsart}

\usepackage{amssymb}

\begin{document}

\begin{frontmatter}



\title{The Wilson-Polchinski exact renormalization group equation}


\author{C.\ Bervillier}
\ead{bervil@spht.saclay.cea.fr}

\address{Service de physique th\'{e}orique,
CEA/DSM/SPhT-CNRS/SPM/URA 2306
 CEA/Saclay,
F-91191 Gif-sur-Yvette C\'{e}dex, France}

\begin{abstract}
The critical exponent $\eta $ is not well accounted for in the
Polchinski exact formulation of the renormalization group (RG). With
a particular emphasis laid on the introduction of the critical
exponent $\eta $, I re-establish (after Golner, hep-th/9801124) the
explicit relation between the early Wilson exact RG equation,
constructed with the incomplete integration as cutoff procedure, and
the formulation with an arbitrary cutoff function proposed later on
by Polchinski. I (re)-do the analysis of the Wilson-Polchinski
equation expanded up to the next to leading order of the derivative
expansion. I finally specify a criterion for choosing the ``best''
value of $\eta $ to this order. This paper will help in using more
systematically the exact RG equation in various studies.
\end{abstract}

\begin{keyword}
Exact renormalization group\sep Derivative expansion \sep Critical
exponents
\PACS 05.10.Cc  \sep 11.10.Gh \sep 64.60.Ak
\end{keyword}
\end{frontmatter}

The renormalization group theory is specifically adapted to treat
physical situations where infinitely many scales are
(continuously) coupled \cite{425}
(on this point see \cite{2839} for example and a recent talk in \cite{5417}%
). In the study of such situations one may expect to have to do a
complicated nonperturbative calculation. No doubt in the spirit of
Wilson: his renormalization group formulation provided us with a
nonperturbative tool to treat a nonperturbative problem \cite{2839}.
However, in practice, the perturbative approach has appeared more
attractive, surprisingly very efficient
---so far as to produce very accurate estimates of
(nonperturbative in nature) universal quantities such as critical
exponents \cite{4608}--- and exposed at length in any text book
dealing with the renormalization group. On the contrary, a
nonperturbative formulation of the renormalization group ---proposed
since 1970 \cite{4495} and named the ``exact'' renormalization group
equation \cite{440} (ERGE) to distinguish it from the discretized
version--- has little been studied before the ninety's
 \cite{4595} due to its disheartening complexity (an
integro-differential equation whose study requires numerical
approximations or truncations not very well controlled). Even after
fifteen years of resurgence of interest in the ERGE \cite{4595},
there is, apparently, little success and when there are new
interesting results \cite{4716}, it seems difficult to establish
their reliability because we are usually not familiar with
nonperturbative approaches.

Contrary to the generally admitted idea \cite[p. 588]{4948} I do
not think that the calculations done with the ERGE are less
precise than those obtained in using perturbative approaches. In
fact one must compare what is effectively comparable. One easily
forgets that the success of perturbation theory in estimating
critical exponents of the Ising and $n$ -vector models is based on
a tremendous numerical tour de force which has provided us with
seven orders of the perturbative series \cite{323} combined with
clever resummation techniques of divergent series, techniques
themselves based on highly nontrivial estimations of large order
behavior of these series \cite {4608,5428}. In summary, this
outstanding success is exceptional and one surely not be able to
reproduce it every day in any other domain of physical interest
(however see  \cite{5702}). It is possible that studies of the
ERGE be never so developed to reach such a success. But if, more
reasonably, one compares the results of the ERGE to those obtained
with the most commonly used perturbation technique, namely the
$\epsilon $-expansion up to, say at most $O\left( \epsilon
^{2}\right) $ then quantitatively the comparison is far from being
ridiculous (see the end of the paper). At the qualitative level,
one should even be able, sometimes in studying an ERGE, to show
that the perturbative approach has failed (or will fail) in
treating correctly some specific physical situation. Let me remind
the reader with a remark relative to the use of the ERGE:
...[perhaps one day]{\sl \ ``one will be able to develop
approximate forms of the transformation which can be integrated
numerically; if so one might be able to solve problems which
cannot be solved any other way}''  \cite[p. 153]{440}. Actually,
as I show in an other paper \cite{5745}, the so-called Lifshitz
tricritical point effectively involves a marginally relevant
coupling which defies the perturbative framework.

With a view to consider correctly the ERGE adapted to the study of
the Lifshitz point \cite{5745} (which involves two correlation
lengths instead of one for the ordinary critical point), it is
necessary to come back to the derivation of the ERGE on one aspect:
the introduction of the critical exponent $\eta $. Actually a
misleading account of this parameter exists since the famous paper
by Wegner and Houghton \cite{414} and has been repeated by the
authors (including those of \cite{4595}) who have tried to determine
(or to report on) the value of $\eta $ from the study of the
Polchinski ERGE \cite{3491,3836}. On the contrary, the study of
Golner \cite {212} who considered the original Wilson version has
been correctly done. As for studies using the ERGE satisfied by the
Legendre transformed action (or the one-particle-irreducible vertex
function) as in \cite{3357} or in \cite {3828} they are basically
correct although the introduction of $\eta $ is effectuated by ad
hoc dimensional arguments which, in my opinion, circumvent the
crucial point of the many scales involved in the problem. The same
kind of appreciation goes also for the calculations done along the
line of the effective average action \cite{3642} initiated by
Wetterich
 \cite{4427} (for a review see \cite{4700}) and for the
calculations done with a proper time regulator \cite{4858}. In the
following I will not consider further these ERGE's since the
Wilson-Polchinski equation is much simpler and allows as well a
correct investigation of the fixed point properties and of the
associated critical exponents what is sufficient to illustrate my
purposes.

Now, let me show the origin of the difficulty.

One currently presents the RG transformation of an action $S\left[
\phi \right] $ as consisting of two steps:

\begin{enumerate}
\item  an integration of the high momentum components of the field
generating an effective action with a reduced cutoff.

\item  a rescaling of the momenta back to the initial value of the cutoff
accompanied by a re-normalization of the field $\phi \rightarrow
\zeta \phi $ ($\zeta $ being related to $\eta $).
\end{enumerate}

In this view the two steps are well separated (besides only the
step 1 is considered in \cite{354}) and it is a matter of fact
that in considering the construction of the ERGE one never makes
reference to any momentum scale other than the current scale
$\Lambda $, except Golner \cite{3912}. In particular, one does not
keep track of the history of the scale changes starting from some
initial momentum scale $\Lambda _{0}$; the whole procedure is
usually done ``instantaneously'' \cite{note2}.

However, in doing so, one forgets two (closely related) Wilson's
prescriptions:

\begin{description}
\item[a)]  the re-normalization of the field ($\zeta $) must be chosen in
such a way as to keep one term of the action unchanged ({\em since
the beginning)}.

\item[b)]  the final RG equation must not depend explicitly on the
renormalization ``time'' $\ell =\Lambda /\Lambda _{0}$.
\end{description}

Prescription a) clearly imposes a memory of the different changes
of scale starting from some initial scale $\Lambda _{0}$. This
memory is usually not considered though it is inherent to the
physics of many scales under study. Prescription b) is a
consequence of a), it implies that the explicit scale dependence
induced by a) be compensated by an appropriate choice of the
cutoff function. Morris \cite{3357} satisfies this prescription
when he considers a cutoff function $C\left( q,\Lambda \right) $
adapted to the scaling behavior expected at a fixed point but he
writes: $C\left( q,\Lambda \right) =\Lambda ^{\eta
-2}\tilde{C}\left( \frac{q^{2}}{\Lambda ^{2}}\right) $ without any
reference to some $\Lambda _{0}$ which though would have been
necessary from simple dimensional analysis arguments ($\Lambda $
is dimensionfull and $\eta $ is not an integer).

The expected relation of equivalence, simply noted by Morris
\cite{3358}, between the Wilson (with a specific cutoff procedure
named the incomplete integration) and Polchinski (with an
arbitrary cutoff function) versions of the ERGE has been
explicitly established in a recent work by Golner \cite {3912}. It
is, however, useful to re-consider explicitly the derivation of
the Polchinski ERGE including the rescaling procedure.

In order to fully take into account the prescriptions a) and b)
listed above, the starting point is a modified version of
Polchinski's action \cite {354}. Consider an action of the
following form:
\begin{equation}
S\left[ \phi \right] =\frac{1}{2}\int_{q}\phi _{q}P^{-1}(\frac{q^{2}}{%
\Lambda ^{2}},\ell )\phi _{-q}+S_{{\rm int}}\left[ \phi \right]
\label{eq:Polch-Action}
\end{equation}
in which $\ell =\frac{\Lambda \,}{\Lambda _{0}}$ is the current
value of the (running) momentum cutoff $\Lambda $ measured in some
fixed initial momentum scale $\Lambda _{0}$.
$P(\frac{q^{2}}{\Lambda ^{2}},\ell )$ is some dimensionless cutoff
function which depends explicitly on $\ell $, contrary to the
original Polchinski's cutoff function. This dependency makes a
reference to some initial scale, it is inherent to the many scale
problem that one is supposed to consider here. It is shown below
that, near a fixed point of relevance for the study of an ordinary
critical point, the explicit $\ell $-dependence factorizes:
\begin{equation}
P(\frac{q^{2}}{\Lambda ^{2}},\ell )=\ell ^{2\varpi }\tilde{P}(\frac{q^{2}}{%
\Lambda ^{2}})  \label{eq:FormeDeP}
\end{equation}
with $\varpi =1-\frac{\eta }{2}$. Notice then that the factorized $\ell $%
-dependence in (\ref{eq:FormeDeP}) is identical to the $\left(
\frac{\Lambda }{\Lambda _{0}}\right) ^{2-\eta }$ factor introduced
by Golner \cite{3912} in his own cutoff function but from ad hoc
dimensional analysis involving the anomalous dimension that the
field acquires at a fixed point ---see also Morris in \cite{3357}.
In fact there is no need to call any anomalous dimensional
analysis in, it suffices to follow the rules established by
Wilson.

The Polchinski flow equation, which corresponds to the step 1
described above (associated to the partial transformation $
{\mathcal G}_{\rm{tra}}S[\phi ]$ and which reduces infinitesimally
the range of integration of $q$) is
unchanged in its general form \cite{4595} and reads ($\dot{S}=-\Lambda \frac{%
{\rm d}S}{{\rm d}\Lambda }={\mathcal G}_{\rm{tra}}S+{\mathcal
G}_{\rm{dil}}S$):

\begin{equation}
{\mathcal G}_{\rm{tra}}S[\phi ] =\frac{1}{2}\int_{q}\Lambda \frac{\partial P%
}{\partial \Lambda }\left[ \frac{\delta ^{2}S}{\delta \phi
_{q}\delta \phi
_{-q}}-\frac{\delta S}{\delta \phi _{q}}\frac{\delta S}{\delta \phi _{-q}}%
 +2P^{-1}\phi _{q}\frac{\delta
S}{\delta \phi _{q}}\right] \label{eq:coarse}
\end{equation}
in which expression the field variable $\phi _{q}$ is such that
$0<\left| {\bf q}\right| /\Lambda <1-{\rm d}\ell /\ell $ (i.e. the
rescaling of the above step 2 has not yet been considered). Now
occurs the re-normalization of the field at fixed momentum scale:
\begin{equation}
\phi _{q}=\zeta \left( \ell -{\rm d}\ell \right) \phi _{q}^{\prime
\prime } \label{eq:re-norm1}
\end{equation}

This step is necessary because one must define a new field so as to
keep the original physics unchanged (this is equivalent to the
Kadanoff prescription
 \cite{248}). Notice here the implicit but obliged reference to an
initial
scale of reference $\Lambda _{0}$ (associated to some original physics) via $%
\ell $. Below it is shown that near a fixed point of relevance in
the study of an ordinary critical point, $\zeta \left( \ell
\right) $ must have the following form:
\begin{equation}
\zeta \left( \ell \right) =\ell ^{\varpi }  \label{eq:dzetadel}
\end{equation}

$\zeta \left( \ell \right) $ induces an explicit dependence on $\ell $ in (%
\ref{eq:coarse}) once expressed in terms of the field $\phi
_{q}^{\prime \prime }$:

\begin{equation}
{\mathcal G}_{\rm{tra}}S[\phi ] =\frac{1}{2}\int_{q}\Lambda \frac{\partial P%
}{\partial \Lambda }\left[ \ell ^{-2\varpi }\left( \frac{\delta
S}{\delta
\phi _{q}^{\prime \prime }}\frac{\delta S}{\delta \phi _{-q}^{\prime \prime }%
}-\frac{\delta ^{2}S}{\delta \phi _{q}^{\prime \prime }\delta \phi
_{-q}^{\prime \prime }}\right) -2P^{-1}\phi _{q}^{\prime \prime
}\frac{\delta S}{\delta \phi _{q}^{\prime \prime }}\right]
\label{eq:Gtra}
\end{equation}
Notice \cite{note3} that since (\ref{eq:coarse}) is already of order ${\rm d}%
\ell $ one may neglect the correction proportional to ${\rm d}\ell
$ in (\ref {eq:re-norm1}).

Requiring that the final equation must not depend explicitly on
$\ell $ (see \cite[p. 126]{440}), then the anticipated $\ell
$-dependence of the cutoff function $P$ is linked to that of
$\zeta $ and using (\ref{eq:dzetadel}) one obtains
(\ref{eq:FormeDeP}).

Finally the rescaling is performed and gives $ {\mathcal
G}_{\rm{dil}}S$ which must globally account for the transformation
of $S$ under the change $\phi _{q}\rightarrow \phi _{q^{\prime
}}^{\prime }$ where $q^{\prime }$ is rescaled back as $q^{\prime
}=(1+\frac{{\rm d}\ell }{\ell })q$ (here I consider dimensionless
$q$). This means that ${\mathcal G}_{\rm{dil}}S$ includes also the
re-normalization of the field given in (\ref{eq:re-norm1}).

Consequently there are two sources of re-normalisation of the field in $%
{\mathcal G}_{\rm{dil}}S$: one coming from $\zeta \left( \ell -{\rm
d}\ell \right) $ and the other from the actual rescaling to the
original momentum scale which expresses as follows:
\begin{equation}
\phi _{q}^{\prime \prime }=s^{y}\phi _{sq=q^{\prime }}^{\prime }
\end{equation}
where$s=1+\frac{%
{\rm d}\ell }{\ell }$ and $y$ is simply determined by usual
dimensional analysis and has already
been implicitly fixed by choosing a dimensionless cutoff function $P$ in (%
\ref{eq:Polch-Action}). This yields $y=\frac{d}{2}$ and thus $\phi
_{q}=\left( 1-\frac{{\rm d}\ell }{\ell }\right) ^{\varpi }\left( 1+\frac{%
{\rm d}\ell }{\ell }\right) ^{\frac{d}{2}}\ell ^{\varpi }\phi
_{q^{\prime }}^{\prime }$ which gives:
\begin{equation}
{\mathcal G}_{\rm{dil}}S=\int_{q}\left[ \left( \frac{d}{2}-\varpi
\right) \phi _{q}+{\bf q}\cdot \partial _{q}\phi _{q}\right]
\,\frac{\delta S}{\delta \phi _{q}}  \label{eq:Gdil}
\end{equation}
in which ${\bf q}\cdot \partial _{q}=\sum_{\mu =1}^{d}q_{\mu
}\frac{\partial }{\partial q_{\mu }}$

By adding up (\ref{eq:Gtra}) and (\ref{eq:Gdil}), the complete
Polchinski
equation finally reads (as already said above, the explicit $\Lambda $%
-dependence in $\tilde{P}(\frac{q^{2}}{\Lambda ^{2}})$ has been
absorbed in a dimensionless $q$):
\begin{eqnarray}
\dot{S} &=&\int_{q}\left[ \left( \frac{d}{2}-\varpi \right) \phi _{q}+{\bf q}%
\cdot \partial _{q}\phi _{q}\right] \,\frac{\delta S}{\delta \phi
_{q}} \nonumber \\
&&+\int_{q}\left[ \varpi \tilde{P}\left( q^{2}\right) -q^{2}\tilde{P}%
^{\prime }\left( q^{2}\right) \right] \left( \frac{\delta
^{2}S}{\delta \phi
_{q}\delta \phi _{-q}}-\frac{\delta S}{\delta \phi _{q}}\frac{\delta S}{%
\delta \phi _{-q}} +2\tilde{P}^{-1}\phi _{q}\frac{\delta S}{\delta
\phi _{q}}\right) \label{eq:PolchNew}
\end{eqnarray}
in which $\tilde{P}\left( x\right) $ is defined in (\ref{eq:FormeDeP}) and $%
\tilde{P}^{\prime }\left( x\right) =\frac{{\rm d}\tilde{P}\left( x\right) }{%
{\rm d}x}$.

Compared to the Wilson equation:

\begin{eqnarray}
\dot{S} &=&\int_{q}\left[ \left( \frac{d}{2}\,\phi _{q}+{\bf
q}\cdot
\partial _{q}\phi _{q}\right) \right] \,\frac{\delta S}{\delta
\phi _{q}}  \nonumber
\\
&&+\int_{q}\left( c+2q^{2}\right) \left( \frac{\delta
^{2}S}{\delta \phi
_{q}\delta \phi _{-q}}-\frac{\delta S}{\delta \phi _{q}}\frac{\delta S}{%
\delta \phi _{-q}} +\phi _{q}\frac{\delta S}{\delta \phi
_{q}}\right)
\end{eqnarray}
it is easy to show that, provided $\varpi =c$, one obtains
\cite{3912} Wilson from Polchinski by choosing $\tilde{P}\left(
q^{2}\right) ={\rm e}^{-2q^{2}}$
and by making the redundant change $\phi _{q}\rightarrow {\rm e}%
^{-q^{2}}\phi _{q}$.

The relation of $\varpi $ to the critical exponent $\eta $ is
obtained from eqs (\ref{eq:re-norm1},\ref{eq:dzetadel}) which
induce a redefinition of the field with a view to keep the physics
unchanged compared to the physics described by the original field.
Notice that this re-normalization of $\phi $ occurs before the
rescaling of the momenta has been performed. We are in position to
use the trivial changes of \cite[p. 592]{301} under the exact form
of equations 2.22 and 2.23 (see also \cite{2727}). Let me consider
the two-point correlation function in terms of the original field:
\begin{equation}
\left\langle \phi _{p}\phi _{-p}\right\rangle _{S}=G(p,S)
\end{equation}

Provided one is only interested to momenta $q=\ell p$ (with $\ell
<1$) then
one has [using (\ref{eq:re-norm1}) with $\ell -{\rm d}\ell \rightarrow \ell $%
]:
\begin{eqnarray}
\phi _{q} &=&\zeta \left( \ell \right) \phi _{q}^{\prime \prime }
\\ \left\langle \phi _{p}\phi _{-p}\right\rangle _{S} &=&\left[
\zeta \left( \ell \right) \right] ^{2}\left\langle \phi
_{q}^{\prime \prime }\phi _{-q}^{\prime \prime }\right\rangle
_{S^{\prime }}
\end{eqnarray}
where $S=S\left[ \phi \right] $ and $S^{\prime }=S^{\prime }\left[
\phi ^{\prime \prime }\right] .$ Finally it comes:
\begin{equation}
G(p,S)=\left[ \zeta \left( \ell \right) \right] ^{2}G(\ell
p,S^{\prime }) \label{eq:Correl2}
\end{equation}
which is a strong constraint on $G$. Indeed, following Fisher
\cite{2727}, one may easily show that at a fixed point $S^{*}$,
$G^{*}(p)$ and $\zeta \left( \ell \right) $ have the respective
following forms:
\begin{eqnarray}
G^{*}(p) &=&\frac{D}{p^{\vartheta }}  \label{eq:2pt} \\ \zeta
\left( \ell \right) &=&\ell ^{\frac{\vartheta }{2}}
\label{eq:dzeta2l}
\end{eqnarray}
with $\vartheta =2\zeta ^{\prime }\left( 1\right) $ and $\zeta
^{\prime }\left( \ell \right) =\frac{{\rm d}\zeta }{{\rm d}\ell
}$. Consequently (\ref {eq:dzetadel},\ref{eq:dzeta2l}) give the
relation $\varpi =\vartheta /2$ and finally since (\ref{eq:2pt})
is precisely the scaling behavior expected for
the two point correlation function at an ordinary critical point with $%
\vartheta =2-\eta $, then one obtains the relation we were looking
for:
\begin{equation}
\varpi =1-\frac{\eta }{2}  \label{eq:eta}
\end{equation}

This ends the derivation of the Wilson-Polchinski ERGE.

Since the previous studies of the Polchinski equation up to
$O\left(
\partial ^{2}\right) $ in the derivative expansion \cite{3491,3836} have not
been done correctly [the term proportional to $\varpi
\tilde{P}\left( q^{2}\right)$ in (\ref{eq:PolchNew}) was not
considered], I find it useful (despite Golner's study \cite{212}) to
reconsider it at the light of the established equivalence between
the Wilson and Polchinski versions.

I consider eq. (\ref{eq:PolchNew}) expanded up to the second
derivative of the field. This means that the action is limited to
the form:
\begin{equation}
S\left[ \phi \right] =\int {\rm d}x\left[ Z(\phi )\left( \partial
\phi \right) ^{2}+V\left( \phi \right) \right]
\label{eq:ActionO2}
\end{equation}
in which $V\left( \phi \right) $ and $Z(\phi )$ are two arbitrary
fonctionnals of $\phi \left( x\right) $. Before writing down the
equations
satisfied by $V\left( \phi \right) $ and $Z(\phi )$ as consequence of eq. (%
\ref{eq:PolchNew}) truncated to actions of type
(\ref{eq:ActionO2}), let me introduce two modifications. First, in
order to allow a comparison with the work of Golner \cite{212}, I
considers a redundant transformation of the field:
\[
\tilde{\phi}_{q}=\psi \left( q^{2}\right) \phi _{q}
\]
in which $\psi \left( q^{2}\right) $ is arbitrary except the normalization $%
\psi \left( 0\right) =1$. Secondly, for practical reason, I
subtract the high temperature fixed point
$\frac{1}{2}\int_{p}\left( \tilde{P}\psi ^{2}\right) ^{-1}\phi
_{p}\phi _{-p}$ from the action. So that in terms of
the new action and the new field (again denoted respectiveley $S$ and $\phi $%
), eq. (\ref{eq:PolchNew}) reads:
\begin{eqnarray}
\dot{S} &=&-\int_{q}\phi _{q}\left( \tilde{d}_{\phi
}\,+2q^{2}\frac{\psi ^{\prime }}{\psi }+{\bf q}\cdot \partial
_{q}\right) \frac{\delta S}{\delta \phi _{q}}\nonumber
\\ &&+\int_{q}\left( \varpi \tilde{P}-q^{2}\tilde{P}^{\prime }\right)
\psi
^{2}\left[ \frac{\delta ^{2}S}{\delta \phi _{q}\delta \phi _{-q}}-\frac{%
\delta S}{\delta \phi _{q}}\frac{\delta S}{\delta \phi
_{-q}}\right]
\end{eqnarray}
in which $\tilde{d}_{\phi }=\frac{d}{2}+\varpi $ and $\psi ^{\prime }={\rm d}%
\psi /{\rm d}q^{2}$.

Two other useful modifications are introduced in the derivative
expansion.
First I rescale the field $\phi =I_{0}^{1/2}\bar{\phi}$ and the potential $%
V=I_{0}\bar{V}$ where $I_{0}=\int_{q}\left( \varpi \tilde{P}-q^{2}\tilde{P}%
^{\prime }\right) \psi ^{2}$ and instead of $V$, I considers the
equation for ${\rm v}_{1}={\rm d}\bar{V}/{\rm d}\bar{\phi}$.
Finally, for the sake of unified notations I rename $Z$ as ${\rm
v}_{2}$. With these new definitions, the RG equations $O(\partial
^{2})$ read:

\begin{eqnarray}
{\rm \dot{v}}_{1} &=&{\rm v}_{1}^{\prime \prime }+d_{\phi }{\rm v}%
_{1}-\left( \tilde{d}_{\phi }\bar{\phi}+2\varpi {\rm v}_{1}\right) {\rm v}%
_{1}^{\prime }+P_{1}{\rm v}_{2}^{\prime }  \label{eq:1} \\ {\rm
\dot{v}}_{2} &=&{\rm v}_{2}^{\prime \prime }-2\left( \varpi
+1\right) {\rm v}_{2}-\left( \tilde{d}_{\phi }\bar{\phi}+2\varpi
{\rm v}_{1}\right) {\rm v}_{2}^{\prime } +{\rm v}_{1}^{\prime
}\left( P_{2}{\rm v}_{1}^{\prime }-2\psi _{0}^{\prime }-4\varpi
{\rm v}_{2}\right)  \label{eq:2}
\end{eqnarray}
in which the prime means the derivative with respect to $\bar{\phi}$ and
 $d_{\phi }=\frac{d}{2}-\varpi $, $\psi _{0}^{\prime }=\left. {\rm d}%
\psi /{\rm d}q^{2}\right| _{q=0}$, $P_{1}=2\frac{I_{1}}{I_{0}}$ with $%
I_{1}=\int_{q}q^{2}\left( \varpi \tilde{P}-q^{2}\tilde{P}^{\prime
}\right) \psi ^{2}$ and $P_{2}=-\left[ \tilde{P}_{0}^{\prime }\left(
\varpi -1\right) +2\varpi \psi _{0}^{\prime }\right] $ with
$\tilde{P}_{0}^{\prime }=\left. {\rm d}%
\tilde{P} /{\rm d}q^{2}\right| _{q=0}$. In order to numerically
study this set of second order differential equations I make the
following choice:
\begin{eqnarray}
\tilde{P}\left( q^{2}\right) &=&{\rm e}^{-aq^{2}} \\ \psi \left(
q^{2}\right) &=&\frac{1}{1+bq^{2}}
\end{eqnarray}
with $a$ and $b$ two parameters on which the value of the critical
exponents do not theoretically depend according to the so-called
(renormalization) scheme independence ($a$) and reparametrization
invariance ($b$). But since these two general properties are no
longer satisfied in the derivative expansion, I shall obtain values
of the critical exponents that depend on $a$ and $b$. The main
question will then be to determine the values of these two
parameters which give the best value of $\eta$ to the order
considered.

The study of eqs (\ref{eq:1},\ref{eq:2}) is by now standard
\cite{3817}. The fixed point equations ${\rm \dot{v}}_{1}={\rm
\dot{v}}_{2}=0$ are two coupled differential equations of second
order for two ordinary functions of the real variable
$\bar{\phi}$. These equations may be numerically integrated using
the shooting method with the Newton-Raphson algorithm (see,
for example, \cite{4394}): starting from a sufficiently large value $\bar{%
\phi}_{0}$ of $\bar{\phi}$ where the regular large $\bar{\phi}$
behavior of the ${\rm v}_{i}$'s is imposed, one shoots towards the
origin $\bar{\phi}=0$ where one checks for even (if one is looking
for a ${\rm Z}_{2}$-symmetric fixed point, odd otherwise)
functions $V$ and $Z$.

The large $\bar{\phi}$ behavior of the solutions corresponding to
eqs (\ref {eq:1},\ref{eq:2}) looks like \cite{note4}:

\begin{eqnarray}
{\rm v}_{1{\rm asy}} &=&G_{1}\bar{\phi}^{\theta _{1}}+\theta _{1}G_{1}^{2}%
\bar{\phi}^{2\theta _{1}-1}+\cdots \\ {\rm v}_{2{\rm asy}}
&=&b\theta _{1}G_{1}\bar{\phi}^{\theta _{1}-1}+\cdots
+G_{2}\bar{\phi}^{\theta _{2}}+\cdots
\end{eqnarray}
with $\theta _{1}=\frac{d-2\varpi }{d+2\varpi }$ and $\theta _{2}=-4\frac{%
1+\varpi }{d+2\varpi }$; $G_{1}$ and $G_{2}$ must be adjusted at $\bar{\phi}=%
\bar{\phi}_{0}$ in order to reach the origin with the following
conditions satisfied (for a ${\rm Z}_{2}$-symmetric fixed point):
\begin{eqnarray}
{\rm v}_{1}(0) &=&0  \label{eq:FPcond1} \\ {\rm v}_{2}^{\prime
}(0) &=&0  \label{eq:FPcond2}
\end{eqnarray}

Since eqs (\ref{eq:1},\ref{eq:2}) are of second order the general
solution involves four arbitrary constants which are fixed by
$G_{1}$, $G_{2}$ and the two conditions
(\ref{eq:FPcond1},\ref{eq:FPcond2}). As for the determination of
$\eta =2\left( 1-\varpi \right) $, it is associated with the
additional condition:
\begin{equation}
{\rm v}_{2}(0)=Z_{0}  \label{eq:Z0cond}
\end{equation}
in which $Z_{0}$ is some arbitrarily fixed constant (the value of
$Z\left( 0\right) $). In principle, $\eta $ should not vary with
$Z_{0}$, but since the reparametrization invariance is broken by
the derivative expansion, one will obtain a non trivial function
$\eta \left( Z_{0}\right) $ which, however, should show, as
vestige of the invariance, an extremum to which is associated a
zero eigenvalue \cite{4420} (see below). The value of $\eta $ at
this extremum ($\eta ^{{\rm opt}}$) being its optimized best value
at the order of the derivative expansion considered (for $a$ and
$b$ fixed).

As usual, I find only one fixed point satisfying the ${\rm
Z}_{2}$-symmetry, it is characterized by a value of $G_{1}$ which is
close to that obtained at the leading order of the derivative
expansion: $G_{1}\simeq -2.3$ (which now depends on the value of
$Z_{0}$). Below, I explain how I have practically chosen the optimum
value of $Z_{0}$.

Once the fixed point has been determined (depending on $a$ and
$b$), one
looks at the eigenvalue equations obtained by linearizing eqs (\ref{eq:1},%
\ref{eq:2}) about a fixed point solution ${\rm v}_{i}^{*}$. By setting ${\rm %
v}_{i}={\rm v}_{i}^{*}+\varepsilon {\rm e}^{\lambda t}{\rm g}_{i}$ (with $%
t=-\ln \ell $) and retaining the linear term in $\varepsilon $,
the eigenvalue equations read:
\begin{eqnarray}
{\rm g}_{1}^{\prime \prime } &=&\left( \lambda -d_{\phi }+2\varpi {\rm v}%
_{1}^{*\prime }\right) {\rm g}_{1}+\left( \tilde{d}_{\phi }\bar{\phi}%
+2\varpi {\rm v}_{1}^{*}\right) {\rm g}_{1}^{\prime }-P_{1}{\rm g}%
_{2}^{\prime }  \label{eq:VP1} \\ {\rm g}_{2}^{\prime \prime }
&=&\left[ \lambda +2\left( \varpi +1+2\varpi
{\rm v}_{1}^{*\prime }\right) \right] {\rm g}_{2}+\left( \tilde{d}_{\phi }%
\bar{\phi}+2\varpi {\rm v}_{1}^{*}\right) {\rm g}_{2}^{\prime
}+2\varpi {\rm v}_{2}^{*\prime }{\rm g}_{1}
 \nonumber \\
&&+2\left( 2\varpi {\rm v}%
_{2}^{*}-P_{2}{\rm v}_{1}^{*\prime }-b\right) {\rm g}_{1}^{\prime
}\label{eq:VP2}
\end{eqnarray}
the interesting solutions of which are again looked for by the
shooting method such that they satisfy the following large
$\bar{\phi}$ behavior at some initial value $\bar{\phi}_{0}$ (with
$\varkappa _{1}=\frac{d-2\varpi
-2\lambda }{d+2\varpi }$, $\varkappa _{2}=-2\frac{2+2\varpi +\lambda }{%
d+2\varpi }$):
\begin{eqnarray}
{\rm g}_{1asy} &=&S_{1}\bar{\phi}^{\varkappa _{1}}+\cdots \\
{\rm g}_{2asy} &=&bS_{1}\varkappa _{1}\bar{\phi}^{\varkappa _{1}-1}+S_{2}%
\bar{\phi}^{\varkappa _{2}}+\cdots
\end{eqnarray}
and the symmetric property at the origin $\bar{\phi}=0$:
\begin{eqnarray}
{\rm g}_{1}\left( 0\right) &=&0  \label{eq:VPsymCond1}
\\ {\rm g}_{2}^{\prime }\left( 0\right) &=&0  \label{eq:VPsymCond2}
\end{eqnarray}

One could think that the four arbitrary constants of integration are
all fixed by the determination of $S_{1}$, $S_{2}$ and the two
conditions (\ref{eq:VPsymCond1},\ref {eq:VPsymCond2}) leaving no
room to any specific determination of the eigenvalue $\lambda $. In
fact, as in any eigenvalue problem, the global normalization of the
eigenvectors may be chosen at will. For example, one may arbitrarily
fix $S_{1}=1$ and this allows the determination of quantized values
for $\lambda $ \cite{3836}.

With the ERGE, one may also easily considers the
asymmetric case (corresponding to an action $S$ involving odd powers of $%
\phi $). To this end it is sufficient to modify eqs
(\ref{eq:VPsymCond1},\ref {eq:VPsymCond2}) into:
\begin{eqnarray}
{\rm g}_{1}^{\prime }\left( 0\right) &=&0  \label{eq:VPasymCond1}
\\ {\rm g}_{2}\left( 0\right) &=&0  \label{eq:VPasymCond2}
\end{eqnarray}

As mentioned above, the optimized value $\eta ^{{\rm opt}}$
associated with a fixed point that satisfies the reparametrization
invariance is associated
to a zero eigenvalue \cite{4420}. I have used this property to determine $%
\eta ^{{\rm opt}}$ directly from a set of differential equations
instead of
determining it by a numerical estimation of a derivative with respect to $%
Z_{0}$. The procedure consists in considering the two fixed point equations (%
\ref{eq:1},\ref{eq:2}) with ${\rm \dot{v}}_{1}={\rm
\dot{v}}_{2}=0$ together with the two eigenvalue equations
(\ref{eq:VP1},\ref{eq:VP2}) with $\lambda $ fixed to zero while
the condition (\ref{eq:Z0cond}) is removed.

As in the Golner study \cite{212}, I observe a strong dependence
(essentially linear) of $\eta ^{{\rm opt}}$ on the parameter $b$ (of
the redundant function $\psi $). This had led Golner to propose a
criterion of choice based on a global minimization of the function
$Z\left( \phi \right) $ (presently corresponding to ${\rm v}_{2}$).
With regard to the few order of the derivative expansion considered
here, there is no unquestionable criterion of choice, simply the
Golner criterion seems reasonable \cite{Litim}. However it appears
that with the second parameter $a$ (of the cutoff function) one can
make $Z\left( \phi \right) $ globally as small as one wants for any
value of $b$ (and thus for essentially any value of $\eta ^{{\rm
opt}}$). Fortunately, it appears also that at fixed value of $\eta
^{{\rm opt}}$ (fixed value of $b$) the geometrical form of the
function $Z\left( \phi \right) $ is independent of $a$ except in
magnitude (there is some similitude transformation which links the
different curves $Z\left( \phi
\right) $ for a fixed $b$). Hence it suffices to choose a specific form of $%
Z\left( \phi \right) $ to get a unique value of $\eta ^{{\rm
opt}}$ (at a
given $b$ but independent of $a$). I have arbitrarily chosen a family of $%
Z\left( \phi \right) $ such that the first two extrema of the
function
(starting from the origin) are disposed symmetrically with respect to the $%
\phi $-axis. This yields, approximately, the value:
\begin{equation}
\eta ^{{\rm opt}}\simeq 0.025  \label{eq:etafinal}
\end{equation}

From this choice of $\eta $ the determination of the eigenvalues
follows without ambiguity.

In the symmetric case there is a unique positive eigenvalue
$\lambda _{1}$
related to the critical exponent $\nu $ ($\nu =1/\lambda _{1}$), I obtain $%
\nu \simeq 0.60$; the following eigenvalue $\lambda _{2}$ is
negative and is related to the correction exponent $\omega $
($\omega =-\lambda _{2}$), I obtain $\omega \simeq 0.87$. Of
course there is also the zero eigenvalue
corresponding to the definition of $\eta ^{{\rm opt}}$. If instead of $%
V^{\prime }$ I had considered the potential $V$ I would have
obtained the supplementary trivial eigenvalue $\lambda _{0}=d$.

In the asymmetric case I find two positive eigenvalues
$\breve{\lambda}_{1}$
and $\breve{\lambda}_{2}$ corresponding respectively to $\left( d+2-\eta ^{%
{\rm opt}}\right) /2$ (associated with the magnetic-like linear coupling $%
h\phi $) and to $\left( d-2+\eta ^{{\rm opt}}\right) /2$
(associated with a redundant $\phi ^{3}$-like term in $S$). The
first interesting eigenvalue is
the negative one $\breve{\lambda}_{3}$ related to correction exponent $%
\omega _{5}$ ($\omega _{5}=-\breve{\lambda}_{3}$) that controls the
leading correction-to-scaling term due to the asymmetry, I obtain
$\omega _{5}\simeq 1.49$.

The same analysis may be done at the lowest order of the
derivative expansion (local potential approximation), in that case
the study is independent of $a$ and $b$ and one finds (for $\omega
_{5{\rm LPA}}$ see
also \cite{4595,5178}): $\eta _{{\rm LPA}}=0$, $\nu _{{\rm LPA}}=0.64956$, $%
\omega _{{\rm LPA}}=0.65574$, $\omega _{5{\rm LPA}}=1.8867$.

Of course, as indicated in the introduction, the quality of these
results cannot be compared with that of the best estimates
obtained from perturbation theory \cite{4211} for a review see
\cite{4608}: $\nu =0.6304\pm 0.0013$, $\eta =0.0335\pm 0.0025$,
$\omega =0.799\pm 0.011$ (although the present estimates are not
so bad regarding the low order of the expansion).
In fact, a more decent comparison is with the $\epsilon $-expansion up to $%
O\left( \epsilon ^{2}\right) $ (or even up to $O\left( \epsilon
^{3}\right) ) $. One knows that this latter expansion is an
asymptotic series the first two or three terms of which only seem to
converge. For example by setting $\epsilon =1$
at orders 1, 2 and 3 for $\eta $ one obtains respectively $0.$, $0.0185$, $%
0.0372$. The value displayed in (\ref{eq:etafinal}) compares
favourably with these numbers. The question is now to know the
effect of the next order in the derivative expansion. It is
presently under study \cite{progress}.

The case of $\omega _{5}$ plainly illustrates my purpose because
the best perturbative estimate has only been done up to $O\left(
\epsilon ^{3}\right) $ \cite{3437} and it is almost impossible to
get a true estimate from the series as shown by the sequence:
2.83, 0.72, 7.36 (the near diagonal Pad\'{e} approximants yield
instead the sequence: 2.83, 1.85, 2.32) compared to 1.89, 1.49 in
the present study, results obtained very easily by simply changing
the conditions of integration of the equations. One sees that even
if no error estimate can still be given, the derivative expansion
up to $O\left(
\partial ^{2}\right) $ compares favourably with the $\epsilon $-expansion up
to $O\left( \epsilon ^{2}\right) $, even up to $O\left( \epsilon
^{3}\right) $.

{\bf Acknowledgements} I am indebted to C. Bagnuls for discussions
and encouragements all along this work.


\begin{thebibliography}{00}

\bibitem{425}  K. G. Wilson, Rev. Mod. Phys. {\bf 47}, 773
(1975).

\bibitem{2839}  K. G. Wilson, in Phase Transitions and Critical Phenomena Vol.
{\bf VI}, p. 1, {\em Ed. by} C. Domb and M.S. Green (Acad. Press,
N.-Y., 1976).

\bibitem{5417}  J. Zinn-Justin, in Progress in Mathematical
Physics Vol. {\bf 30}, p. 55, {\em Ed. by} B. Duplantier and V.
Rivasseau (Birkhauser-Boston, 2002).

\bibitem{4608}  J. Zinn-Justin, Phys. Rep. {\bf %
344}, 159 (2001).

\bibitem{4495}  K. G. Wilson, in ``{\em Irvine Conference (unpublished)}''
(1970).

\bibitem{440}  K. G. Wilson and J. Kogut, Phys. Rep. {\bf 12C}, 77 (1974).

\bibitem{4595}  C. Bagnuls and C. Bervillier, Phys. Rep. {\bf 348},
91 (2001).

\bibitem{4716}  M. Tissier, B. Delamotte and D. Mouhanna, Phys. Rev. Lett. {\bf 84}%
, 5208 (2000); O. Lauscher and M. Reuter, Phys. Rev. {\bf D65},
025013 (2001).

\bibitem{4948}  A. Pelissetto and E. Vicari, Phys. Rep. {\bf 368}, 549 (2002).

\bibitem{323}  B. G. Nickel, D. I. Meiron and G. A. Baker, Jr, ``{\em %
Compilation of 2-pt and 4-pt graphs for continuous spin models}'',
Guelph University preprint, unpublished (1977).

\bibitem{5428}  F. Jasch and H. Kleinert, J. Math. Phys. {\bf 42}, 52 (2001).

\bibitem{5702}  H. Kleinert and V. I. Yukalov, cond-mat/0402163.

\bibitem{5745}  C. Bervillier, hep-th/0405027.

\bibitem{414}  F. J. Wegner and A. Houghton, Phys. Rev. {\bf A8}, 401
(1973).

\bibitem{3491}  R. D. Ball, P. E. Haagensen, J. I. Latorre and E. Moreno,
Phys. Lett. {\bf B347}, 80 (1995).

\bibitem{3836}  J. Comellas, Nucl. Phys. {\bf
B509}, 662 (1998).

\bibitem{212}  G. R. Golner, Phys. Rev. {\bf B33},
7863 (1986).

\bibitem{3357}  T. R. Morris, Phys. Lett. {\bf B329}, 241 (1994).

\bibitem{3828}  T. R. Morris and M. D. Turner, Nucl.
Phys. {\bf B509}, 637 (1998).

\bibitem{3642}  N. Tetradis and C. Wetterich, Nucl. Phys. {\bf B422}, 541
(1994). L. Canet, B. Delamotte, D. Mouhanna and J. Vidal, Phys. Rev.
{\bf D67}, 065004 (2003); {\em ibid.}{\bf B68}, 064421 (2003). H.
Ballhausen, hep-th/0303070.

\bibitem{4427}  C. Wetterich, Nucl. Phys. {\bf B352}, 529 (1991).

\bibitem{4700}  J. Berges, N. Tetradis and C. Wetterich, Phys. Rep. {\bf 363}, 223 (2002).

\bibitem{4858}  A. Bonanno and D. Zappal\`{a}, Phys. Lett. {\bf B504}, 181 (2001).
 M. Mazza and D. Zappal\`{a}, Phys. Rev. {\bf D64}, 105013
(2001).

\bibitem{354}  J. Polchinski, Nucl. Phys. {\bf B231}, 269 (1984).

\bibitem{3912}  G. R. Golner, hep-th/9801124.

\bibitem{note2}  When the one-particle-irreducible vertex function is
considered, two momentum scales of reference coexist for purely
technical reasons: the current cutoff $\Lambda $ is said to be an
``infrared cutoff'' and the customary ``ultraviolet cutoff''
$\Lambda _{0}$ is still mentioned but plays actually no role. With
regards to the considerations of the present paper, the reader could
be deceived by this apparent memory of the scale changes.

\bibitem{3358}  T. R. Morris, Phys. Lett. {\bf B334}, 355 (1994).

\bibitem{248}  L. P. Kadanoff, Physics {\bf 2}, 263 (1966).

\bibitem{note3}  Notice also that in \cite{414} $\zeta \left( \ell \right) $
which is written $\zeta \left( 1-dt\right) $ is not considered as
contributing to ${\mathcal G}_{\rm{tra}}S[\phi ]$ because precisely
it would induce a term of higher order in $dt$. This argument could
have been correct if one had not the prescription a) which imposes
to keep exactly constant one term of the action. In addition, with a
hard cutoff, prescription b) would have been difficult to realize.

\bibitem{301}  S. K. Ma, Rev. Mod. Phys. {\bf 45}, 589 (1973).

\bibitem{2727}  M. E. Fisher, in ``{\em Critical Phenomena}'',
Lecture Notes in Physics, p. 1, {\em Ed. by} F.J.W. Hahne
(Springer-Verlag, Pub., 1983).

\bibitem{3817}  T. R. Morris, Phys. Rev. Lett. {\bf 77}, 1658 (1996).

\bibitem{4394} W. H. Press, B. P. Flannery, S. A. Teukolsky and W.
T. Vetterling, in ``{\em Numerical Recipes.}'', ``{\em The Art of
Scientific Computing}'' (Cambridge University Press, 1986).

\bibitem{note4}  More terms are needed to numerically obtain a solution with
some accuracy.

\bibitem{4420}  T. L. Bell and K. G. Wilson, Phys. Rev. {\bf B10}, 3935 (1974).

\bibitem{5178}  M. M. Tsypin,
Nucl. Phys. {\bf B636}, 601 (2002).
  D. F. Litim and L. Vergara,
Phys. Lett. {\bf B581}, 263 (2004).

\bibitem{4211}  R. Guida and J. Zinn-Justin, J. Phys. {\bf A31}, 8103 (1998).

\bibitem{progress}  C. Bagnuls, C. Bervillier and M.\ Shpot, in progress

\bibitem{3437} F. C. Zhang and R. K. P. Zia, J. Phys. {\bf A15}, 3303 (1982).

\bibitem{Litim} Criterion such as elaborated by D. F. Litim,
 Phys. Lett. {\bf B486}, 92 (2000), being
 specific to the Legendre transformed ERGE, seems not applicable here.
\end{thebibliography}
\end{document}